# Stability of $CO_2$ Atmospheres on Desiccated M Dwarf Exoplanets

**Short Title: $CO_2$ in M Dwarf Exoplanet Atmospheres**

Peter Gao[1,*], Renyu Hu[1,2,3], Tyler D. Robinson[4,5], Cheng Li[1], and Yuk L. Yung[1]

[1]*Division of Geological and Planetary Sciences, California Institute of Technology, Pasadena, CA 91125, USA*

[2]*Jet Propulsion Laboratory, California Institute of Technology, Pasadena, CA 91109, USA*

[3]*Hubble Fellow*

[4]*NASA Postdoctoral Program Fellow, Ames Research Center, Mountain View, CA, USA, 94035*

[5]*Oak Ridge Associated Universities*

*Corresponding author at: *Division of Geological and Planetary Sciences, California Institute of Technology, Pasadena, CA, USA, 91125*

*Email address:* pgao@caltech.edu

*Phone number:* 626-298-9098

**Abstract**

We investigate the chemical stability of $CO_2$-dominated atmospheres of desiccated M dwarf terrestrial exoplanets using a 1-dimensional photochemical model. Around Sun-like stars, $CO_2$ photolysis by Far-UV (FUV) radiation is balanced by recombination reactions that depend on water abundance. Planets orbiting M dwarf stars experience more FUV radiation, and could be depleted in water due to M dwarfs' prolonged, high-luminosity pre-main sequences (Luger & Barnes 2015). We show that, for water-depleted M dwarf terrestrial planets, a catalytic cycle relying on $H_2O_2$ photolysis can maintain a $CO_2$ atmosphere. However, this cycle breaks down for atmospheric hydrogen mixing ratios <1 ppm, resulting in ~40% of the atmospheric $CO_2$ being converted to CO and $O_2$ on a time scale of 1 Myr. The increased $O_2$ abundance leads to high $O_3$ concentrations, the photolysis of which forms another $CO_2$-regenerating catalytic cycle. For atmospheres with <0.1 ppm hydrogen, $CO_2$ is produced directly from the recombination of CO and O. These catalytic cycles place an upper limit of ~50% on the amount of $CO_2$ that can be destroyed via photolysis, which is enough to generate Earth-like abundances of (abiotic) $O_2$ and $O_3$. The conditions that lead to such high oxygen levels could be widespread on planets in the habitable zones of M dwarfs. Discrimination between biological and abiotic $O_2$ and $O_3$ in this case can perhaps be accomplished by noting the lack of water features in the reflectance and emission spectra of these planets, which necessitates observations at wavelengths longer than 0.95 μm.

# 1. INTRODUCTION

M dwarfs are prime targets for the detection and characterization of terrestrial exoplanets by the James Webb Space Telescope, as they are abundant in the solar neighborhood and their small radii allow for greater transit signals from Earth-sized exoplanets (e.g. Quintana et al. 2014). As a result, terrestrial exoplanets in orbit of M dwarfs may offer the best opportunities for the detection of biosignatures in the near future, such as $O_2$ and $O_3$. However, these species are also produced by $CO_2$ photolysis, leading to potential false positives in biosignature detection. Previous studies have shown that low $H_2O$, high $CO_2$ atmospheres tend to generate more $O_2$ and $O_3$, though the conclusions are very sensitive to surface boundary conditions and atmospheric escape rates (Selsis et al. 2002; Segura et al. 2007; Zahnle et al. 2008; Hu et al. 2012; Tian et al. 2014; Domagal-Goldman et al. 2014).

Recent works by Luger & Barnes (2015) and Tian & Ida (2015) have shown that M dwarf terrestrial exoplanets may be severely depleted in water relative to Earth. Prior to the Main Sequence, M dwarfs evolve along the Hayashi Track for 100 Myr – 1 Gyr with much higher luminosities than later in their evolution (Hayashi 1961). Therefore, any planets in the habitable zone of a main sequence M dwarf would experience a prolonged period of high insolation before the main sequence, leading to a Runaway Greenhouse phase and the loss of much of their atmospheric and surficial water. Their crust and upper mantle could also become desiccated if a surface magma ocean forms as a result of the Runaway Greenhouse, in which case they could be heavily oxidized by acting as a sink for the excess oxygen generated by water photolysis and hydrogen escape (Luger & Barnes 2015). Such planets would be vastly different from Earth in terms of atmospheric chemistry and subsequent evolution. For example, the low atmospheric water content, whether as a result of the initial global desiccation or low water outgassing rates due to the desiccation of the mantle, could lead to high abundances of abiotic $O_2$ and $O_3$ in the event that the remaining atmosphere is dominated by outgassed $CO_2$.

The stability of $CO_2$ in terrestrial planet atmospheres depends on a balance between destruction by Far-UV (FUV) radiation ($\lambda \sim 121 - 200$ nm) from their host stars in the upper atmosphere via

$$CO_2 + h\nu \rightarrow CO + O \tag{R1}$$

and regeneration through the following catalytic cycle in the lower atmosphere:

$$H + O_2 + M \rightarrow HO_2 + M \tag{R2}$$
$$O + HO_2 \rightarrow OH + O_2 \tag{R3}$$
$$OH + CO \rightarrow CO_2 + H \tag{R4}$$
$$\text{----------------------------------}$$
$$CO + O \rightarrow CO_2, \tag{S1}$$

where M is a third body, and the $HO_x$ species ( = H + OH + $HO_2$) are derived from $H_2O$, with small contributions from $H_2O_2$ photolysis by Near-UV (NUV) radiation ($\lambda \sim 300 - 400$ nm). This process is usually faster than the spin-forbidden regeneration reaction

$$CO + O + M \rightarrow CO_2 + M. \tag{R5}$$

S1 is currently active on Mars and Venus (McElroy & Donahue 1972; Nair et al. 1994; McElroy et al. 1973), which explains their $CO_2$-dominated atmospheres despite photolysis by Solar FUV. However, M dwarfs produce much lower NUV relative to FUV due to their lower effective temperatures and intense stellar activity (Tian et al. 2014; West et al. 2004), which would reduce the efficiency of S1 while not affecting, or even increasing R1. Furthermore, if the atmospheric water content were low, as is possibly the case for terrestrial exoplanets in the M dwarf habitable zone, then S1 would be reduced even further. Previous studies that investigated $CO_2$ photochemistry either assumed present/young Sun-like host stars (Selsis et al. 2002; Segura et al. 2007; Zahnle et al. 2008; Hu et al. 2012) and/or ocean worlds (Tian et al. 2014; Domagal-Goldman et al. 2014). In this work, we explore the stability of $CO_2$ on a desiccated planet around an M dwarf and consider the potential for false positive detections of $O_2$ and $O_3$ as biosignatures.

In Section 2 we give a brief summary of the 1-dimensional photochemical model we use to obtain our results, as well as describe the construction of our model atmosphere and the specific cases we considered regarding atmospheric hydrogen content. We present our results for the stability of $CO_2$ atmospheres in Section 3. In Section 4 we discuss the possibility for false positive detections of biosignatures, as well as consider the effects of temperature variations, clouds, hydrogen escape, and interactions between the atmosphere and the surface. We offer our conclusions in Section 5.

## 2. METHODS

*2.1. Photochemical and Transport Model*

We use the Caltech/JPL 1-Dimensional Photochemical model (Allen et al. 1981), as modified by Nair et al. (1994) for the Martian atmosphere. The model calculates the steady state distributions of 27 chemical species: O, O($^1$D), $O_2$, $O_3$, N, N($^2$D), $N_2$, $N_2O$, NO, $NO_2$, $NO_3$, $N_2O_5$, $HNO_2$, $HNO_3$, $HO_2NO_2$, H, $H_2$, $H_2O$, OH, $HO_2$, $H_2O_2$, CO, $CO_2$, $O^+$, $O_2^+$, $CO_2^+$, and $CO_2H^+$. We refer the reader to Allen et al. (1981) for a detailed description of the general photochemical model and Table 2 of Nair et al. (1994) for the full list of reactions and associated rate constants (with minor updates) considered in our model. Table 1 gives a subset of Nair et al. (1994)'s reactions that are significant to this work.

Figure 1 shows a comparison between the solar spectrum (WMO 1985), and a spectrum of the M dwarf GJ 436 from France et al. (2013) that we use for our model, scaled such that the total flux of each star is equal to the total solar flux at 1 AU, ~1360 W m$^{-2}$. In addition, we show the wavelengths at which photolysis of $CO_2$, $O_2$, $O_3$, and $H_2O_2$ is most efficient. It is clear that the GJ 436 spectrum features a much higher FUV/NUV ratio than Sun-like stars, and as a result the photolysis rates of the four aforementioned species will differ between the two cases. The entire spectral range used in the model is ~100 to 800 nm. We assume that the spectrum of the star is fixed throughout the duration of our simulations. This ignores the effects of major flaring events or quiet periods, which we assume to be irrelevant on long time scales.

*2.2 Boundary Conditions*

The choice of boundary conditions strongly affects the resulting atmospheric composition. The major processes that are typically modeled include hydrogen escape, volcanic outgassing, and surface deposition of oxidants. The rates of these processes, especially the latter two, are highly uncertain. For example, the outgassing flux of water for Earth is $10^{11} - 10^{12}$ cm$^{-2}$ s$^{-1}$ (Fischer 2008), while on Venus it could be as low as $5.5 \times 10^5 - 10^7$ cm$^{-2}$ s$^{-1}$ (Walker et al. 1970; Grinspoon 1993). This is likely due to the hydrodynamic loss of water in Venus's past (Hamano et al. 2013), similar to the conditions experienced by our modeled M dwarf exoplanet. Likewise, the deposition velocity of oxidants, such as $O_2$ and $O_3$, is strongly tied to surface composition and water content. On Earth and Mars, aqueous reactions of molecular oxygen and ozone with ferrous iron, sulfides, and other reducing species act as significant sinks for oxidants, but deposition velocities on Mars could be $10 - 100$ times lower than on Earth due to the reduced surface humidity and differences in surface composition (Zahnle et al. 2008). Therefore, given that our modeled planet could be drier still, and that the crust and upper mantle may already be heavily oxidized, the deposition of oxidants could be severely limited.

If hydrogen escape is in the diffusion-limited regime, then the flux can be expressed as

$$\phi_{lim} = f_H(z) b_{H/H_2}^{CO_2} \left( \frac{1}{H_{co_2}} - \frac{1}{H_{H/H_2}} \right) \quad (1)$$

(Zahnle et al. 2008) where $\phi_{lim}$ is the diffusion-limited escape flux, $f_H$ is the total mixing ratio of all hydrogen species at the homopause, $b_{H/H_2}^{CO_2}$ is the binary diffusion coefficient of hydrogen (H or $H_2$) escaping through $CO_2$, and $H_x$ is the scale height of species x, where x = H or $H_2$ and $CO_2$. As M dwarfs emit abundant XUV ($\lambda$ < 121 nm) radiation and highly charged particles that can cause hydrodynamic escape and sputtering, the rate of hydrogen loss could readily reach the diffusion-limited regime (Cohen et al. 2015), though sputtering could be reduced if a strong planetary magnetic field exists (Lammer et al. 2007). On the other hand, the abundance of hydrogen above the homopause is strongly tied to the rate of production from photolysis of water and $H_2O_2$, as well as chemistry in the ionosphere (Nair et al. 1994). Thus, in the limit of low atmospheric hydrogen content, the hydrogen escape flux could be much lower than the diffusion-limited flux.

Given the possibility that the outgassing, deposition, and escape fluxes may be low, we will simplify the problem and focus strictly on the effects of high UV flux, low water content, and high $CO_2$-content by assigning zero fluxes to all species for our upper and lower boundary conditions. We make an exception for ions, which have a zero flux upper boundary condition and a fixed concentration of 0 cm$^{-3}$ at the lower boundary to simulate recombination with electrons at the surface (Nair et al. 1994). This is equivalent to hydrogen escape being balanced by the outgassing of water, with the outgassed oxygen atoms being lost through surface deposition of oxidants. These simplified boundary conditions, which are potentially representative of desiccated terrestrial exoplanets orbiting M dwarfs, will be evaluated in section 4.1.

*2.3. Model Atmosphere*

In order to model a $CO_2$-dominated atmosphere we simulate Mars-like atmospheric composition with a surface atmospheric pressure of 1 bar for an Earth-sized planet with Earth's surface gravity. We keep the same (Martian) mixing ratios of atmospheric species as Nair et al. (1994). The pressure $P$, temperature $T$, and eddy diffusion coefficient $K_{zz}$ profiles are shown in Figure 2. The model atmosphere extends from the surface to 100 km altitude with an altitude bin thickness of 1 km. The pressure profile is calculated assuming an ideal gas atmosphere in hydrostatic equilibrium.

The temperature profile is split into the troposphere, the stratosphere/mesosphere, and the thermosphere. The troposphere is assumed to be a dry adiabat with the heat capacity of $CO_2$ as a function of temperature taken from McBride & Gordon (1961) and Woolley (1954). The surface temperature is assumed to be 240 K, accounting for the warming due to 1 bar of $CO_2$ (Kasting 1991). The temperature decreases along the adiabat with increasing altitude until it reaches the planetary skin temperature of 139 K, identical to that of Mars, by assumption (Nair et al. 1994). This places the tropopause of our atmosphere at ~7.5 km. Above the tropopause the temperature profile is assumed to be isothermal, consistent with previous works (Kasting 1990; Segura et al. 2005; Segura et al. 2007). The temperature increases in the thermosphere above 60 km due to inefficient collisional cooling, and the temperatures are calculated by assuming that the temperature-pressure relationship is identical in the thermospheres of both our model atmosphere and Nair et al. (1994)'s model atmosphere. This assumption is valid given the density (and thus pressure) dependence of collisional cooling (López-Puertas & Taylor 2001, pp. 77). As the pressure profile depends on the temperature profile, we iterate between the two until convergence.

The eddy diffusion coefficient as a function of altitude is calculated assuming free convection (Gierasch & Conrath 1985, pp. 121),

$$K_{zz} = \frac{H}{3}\left(\frac{L}{H}\right)^{4/3}\left(\frac{R\sigma T^4}{\mu\rho C_p}\right)^{1/3} \tag{2}$$

where $H$ is the scale height given by $RT/\mu g$, $R$ is the universal gas constant, $\sigma$ is the Stefan-Boltzmann constant, $\mu$ is the atmospheric molecular weight, which is assumed to be identical to that of $CO_2$, $\rho$ is the atmospheric mass density, $C_p$ is the atmospheric heat capacity, and $L$ is the mixing length defined with respect to the convective stability of the atmosphere (Ackerman & Marley 2001),

$$L = H\,max(0.1, \Gamma/\Gamma_a) \tag{3}$$

where $\Gamma$ is the lapse rate and $\Gamma_a$ is the adiabatic lapse rate. As with Ackerman & Marley (2001), we set a lower bound for $L$ of *0.1H* in the radiative regions of the atmosphere. Using equations 1 and 2, we find a surface eddy diffusion coefficient value of $3.4 \times 10^7$ cm$^2$ s$^{-1}$, which is more than twice as high as in previous studies where a value of $10^5$ cm$^2$ s$^{-1}$ is often used (e.g. Nair et al. 1994; Segura et al. 2007). This is likely due to the equations' assumption of free convection breaking down in the presence of a solid surface. Therefore, we scale the entire profile by a constant factor such that the surface value is $10^5$ cm$^2$ s$^{-1}$.

The temperature, pressure, and eddy diffusion coefficient profiles are held constant with time throughout the simulation, even when the atmospheric composition changes drastically. We will discuss how a self-consistent atmospheric model may impact our results in Section 4.3.

*2.4. Atmospheric Hydrogen and Oxygen Content*

The hydrodynamic loss of water leads to an atmosphere depleted in hydrogen, which is mostly bound in $H_2$ and $H_2O$ in our model. To simulate such depletion we decrease the initial water content of the atmosphere such that it is undersaturated at all altitudes. Using the expression of Lindner (1988) for the temperature-dependent saturation vapor pressure of water we find the minimum value of the saturation water vapor mixing ratio of our model atmosphere to be $1.675 \times 10^{-5}$ ppm at the tropopause, where the temperature is 139 K. We then vary the $H_2$ content to investigate the result of different degrees of hydrogen depletion in the atmosphere. We simulate six cases, with total atmospheric H mole fractions *f(H)$_{tot}$* of (1) 26 ppm, (2) 2.6 ppm, (3) 0.27 ppm, (4) 0.032 ppm, (5) 0.0088 ppm, and (6) 0.0065 ppm. *f(H)$_{tot}$* is defined as the number of hydrogen atoms in the atmosphere divided by the total number of atoms in the atmosphere. By comparison, the H mole fraction of Venus is 14 ppm (Miller-Ricci et al. 2009).

We assume that the excess oxygen produced from water depletion is taken up by the crust and upper mantle during the Runaway Greenhouse phase, when a magma ocean may have formed on the surface that would have acted as an extremely efficient oxygen sink (Luger & Barnes; Hamano et al. 2013). The resulting atmosphere would then be low in oxygen, while the crust and upper mantle compensate by being oxidized.

## 3. RESULTS

Despite the low water content of our model atmosphere, >50% of the atmospheric $CO_2$ present at the beginning of our simulations is retained. We find that this results from the actions of $CO_2$-regenerating catalytic cycles that rely on $H_2O_2$ and $O_3$ instead of $H_2O$, which arise due to the low M dwarf NUV flux limiting $H_2O_2$ and $O_3$ destruction by photolysis. Nonetheless, abiotic $O_2$ and $O_3$ column mixing ratios up to 20% and 1 ppm, respectively, can be produced from $CO_2$ photolysis, giving rise to potential false positives when using $O_2$ or $O_3$ as the signature of biological photosynthesis.

Figure 3 shows the steady state mixing ratios of several species in our model atmosphere as a function of altitude after 10 Gyr of evolution. The long-lived species, such as $CO_2$, $CO$, $O_2$, $H_2$, $H_2O$, and $H_2O_2$, tend to be well mixed below the homopause (~50 km in our model atmosphere). Above the homopause, (1) $CO_2$ is depleted due to photolysis, which in turn produces an excess of CO and $O_2$, (2) the $H_2$ mixing ratio increases due to molecular diffusion, and (3) $H_2O$ and $H_2O_2$ are fairly well-mixed, as most of their chemistry takes place at lower altitudes. The short-lived species, as expected, are not well mixed: $O_3$ and $HO_2$ both show clear mixing ratio maxima near the homopause due to their high production rates there, while H and OH are only abundant above the homopause due to their rapid removal below.

Figure 3 also shows a clear trend in the mixing ratios of these species as the atmospheric hydrogen content is varied. For example, as the hydrogen content drops, the hydrogen species' mixing ratios also decrease, while the $O_3$ mixing ratio increases. This trend is better seen in Figure 4, which shows the column-averaged mixing ratios of the same species as in Figure 3, as a function of the atmospheric hydrogen content. There is a clear trend of decreasing $CO_2$ with decreasing atmospheric hydrogen, eventually reaching a lower limit for cases with hydrogen mixing ratio ~1 ppm. The high abiotic $O_2$ and CO concentrations of our atmospheres (5 – 20% and 10 – 40%, respectively) are consistent with previous studies showing the propensity for planets with low atmospheric $H_2O$ abundance to lose more $CO_2$ to photolysis than planets with high atmospheric $H_2O$ abundance (e.g. Segura et al. 2007). However, we do not lose as much $CO_2$ as the results of Zahnle et al. (2008), which showed that an atmosphere as depleted in water as ours (~0.6 precipitable microns for our "wettest" case) should be unstable to conversion to CO if irradiated by a Sun-like star. This suggests the existence of alternate chemical pathways aside from S1 that are responsible for regenerating $CO_2$ in low $H_2O$, initially $CO_2$-dominated atmospheres of terrestrial planets orbiting M dwarfs. These chemical pathways must be such that (1) $CO_2$ still dominates even when $H_2O$ concentrations are low, and (2) $CO_2$ is stopped from complete conversion to CO and $O_2$ even when atmospheric hydrogen is severely depleted.

Figure 5 shows the rates of several key reactions near the surface of the planet as the hydrogen content of the atmosphere is reduced from panels 1 to 6, which correspond to Cases 1 to 6, respectively, as well as the points on the column-averaged mixing ratio curves of Figure 4. As catalytic cycles require one reaction to feed into another, the reactions that make up the cycle must all have the same effective reaction rate, determined by the rate-limiting step. Thus, for Case 1, R2, R4, R6, and R7 form a catalytic cycle below 20 km:

$$2(H + O_2 + M \rightarrow HO_2 + M) \quad (R2)$$
$$2HO_2 \rightarrow H_2O_2 + O_2 \quad (R7)$$
$$H_2O_2 + h\nu \rightarrow 2OH \quad (R6)$$
$$2(OH + CO \rightarrow CO_2 + H) \quad (R4)$$
$$\text{---------------------------------}$$
$$2CO + O_2 \rightarrow 2CO_2. \quad (S2)$$

This cycle depends on relatively high concentrations of $H_2O_2$, which replaces water as the major reservoir/source of $HO_x$ species. The lower panel of Figure 4 shows that the equilibrium $H_2O_2$ mixing ratio for the highest $H_2$ case is ~0.2 ppm, which is ~20 times higher than the modeled Martian $H_2O_2$ mixing ratio from Nair et al. (1994). This buildup of $H_2O_2$ is due to its low photolysis rate around M dwarfs. S2 has been considered as a $CO_2$ regeneration pathway for a wet Mars (Parkinson & Hunten 1972; Yung & DeMore 1999), which relies on high $H_2O_2$ abundances arising from high $H_2O$ abundances.

As the atmospheric hydrogen content is decreased from Cases 1 to 3 in Figure 5, the mixing ratios of $H_2O_2$ and $HO_2$ decrease (see Figure 4), resulting in a drop in the rates of R6 and R7. Consequently, S2 becomes less efficient at regenerating $CO_2$. This leads to the decrease in $CO_2$ and the increase in CO and $O_2$ shown in Figure 4, which in turn results in the increase in $O_3$. The rise of ozone increases the rate of R8 and its own photolysis reaction R9, which produces an abundance of free oxygen atoms that go into increasing the rate of R5. Eventually, R8 replaces R6 and R7 in S2, forming

$$H + O_2 + M \rightarrow HO_2 + M \quad (R2)$$
$$O_3 + HO_2 \rightarrow OH + 2O_2 \quad (R8)$$
$$OH + CO \rightarrow CO_2 + H \quad (R4)$$
$$\text{-----------------------------------}$$
$$CO + O_3 \rightarrow CO_2 + O_2, \quad (S3)$$

which, together with R5, dominates the $CO_2$ regeneration process below 15 km. Case 3 shows the switch from S2 to S3 and R5, where neither process is particularly efficient; this causes the minimum in the $CO_2$ mixing ratio curve in Figure 4. The increase in $O_3$ concentrations goes on to increase the rates of S3 and R5 in Cases 4 to 6, which in turn explains the upswing in the $CO_2$ mixing ratio seen in Figure 4 for the cases with atmospheric hydrogen content <0.1 ppm. The importance of $O_3$ in S3 and R5 creates a negative feedback, where excess $CO_2$ photolysis produces excess $O_2$ and $O_3$, which then increases the intensity of $CO_2$ regeneration via S3 and R5. This stops further decreases in the $CO_2$ mixing ratio with decreasing atmospheric hydrogen content. S3 has been recognized as a major catalytic cycle for the loss of $O_3$ on M dwarf exoplanets by Grenfell et al. (2013), but they considered an atmosphere similar to that of modern Earth instead of a $CO_2$-dominated atmosphere like that of our model.

Further evidence of the importance of $O_3$ to $CO_2$ regeneration in these low atmospheric hydrogen/water cases is given in Figure 6, which shows molecular fluxes from Case 1 (top) and Case 6 (bottom) near the planet surface for $O_2$, $O_3$, CO, and $CO_2$. The fluxes of Case 1 show a balance between downwelling CO and upwelling $CO_2$, consistent with the conversion from CO to $CO_2$ occurring near the surface by S2. The fluxes of Case 6 are similar, but, in addition, $O_2$ and $O_3$ also have a correspondence, with conversion of $O_3$ to $O_2$ occurring near the surface in keeping with S3.

For the cases where the atmospheric hydrogen mixing ratio is <0.05 ppm, as in Cases 5 and 6 of Figure 5, R5 dominates and S3 is reduced, due to low $HO_2$ and OH concentrations.

Further decreases in atmospheric hydrogen content are inconsequential since R5 does not depend on any hydrogen species.

The importance of $H_2O_2$ and $O_3$ photolysis for S2, S3, and R5 means our results would differ significantly for similar planets around Sun-like stars. Aside from the lack of a prolonged high luminosity pre-main sequence phase, Sun-like stars also emit more strongly in the NUV, resulting in higher rates of $H_2O_2$ and $O_3$ photolysis (see Figure 1). Therefore, only a small amount of $H_2O_2$ and $O_3$ is necessary to produce the photolysis products needed to regenerate $CO_2$. An alternate simulation of Case 6 that uses a Solar spectrum rather than an M dwarf spectrum yields ~2% $O_2$ and 4% CO, consistent with the results of previous works that used solar spectra (e.g. Hu et al. 2012). Though these values are lower than our nominal results, they are still ~100 times higher than that of a non-desiccated planet orbiting a Sun-like star, such as Earth during the Proterozoic eon (Planavski et al. 2014). This suggests that false positive detections of biosignatures are possible even for desiccated planets around Sun-like stars. These results are also similar to those of Nair et al. (1994) when they only used R5 for $CO_2$ regeneration, though there is more $CO_2$ remaining at equilibrium in our case due to the higher atmospheric density, which increases the rate of R5.

## 4. DISCUSSION

*4.1. Surface Fluxes and Atmospheric Escape*

We have assumed zero-flux boundary conditions for all neutral species in our model atmosphere in order to simplify the problem and focus on the effects of high UV flux, low water content, and high $CO_2$-content. This is equivalent to assuming that all fluxes are nonzero, but are in balance with each other. In this section we evaluate this assumption by introducing explicit, balanced fluxes into our model with the goal of determining what surface and escape fluxes are necessary to reproduce our zero-flux results, and thus what planets may possess these atmospheric compositions. The fluxes modeled are the outgassing of $H_2O$, the escape of H and $H_2$ into space, and the surface deposition of $O_3$. $O_2$ deposition is assumed to be zero due to its relatively low reactivity and solubility in water compared to $O_3$ (Sonntag & von Gunten 2012), though its high abundance in the drier cases may increase its significance (see below). Comparisons between the zero-flux and explicit flux models are done using our wettest and driest cases, Cases 1 and 6, respectively. All other run parameters of the models are identical to the corresponding zero-flux cases.

Figure 7 shows the comparison between the mixing ratios derived from the explicit flux model (dashed curves) and that derived from our nominal model (solid curves), for Cases 1 (red) and 6 (blue). There is satisfactory agreement between the two models for almost all of the species, including those vital to the catalytic cycles outlined in Section 3. The major difference between the two models is the reservoir of atmosphere hydrogen. For the zero-flux case, the reservoirs are $H_2$ and $H_2O_2$, while in the explicit flux model the hydrogen reservoir is water. This is understandable since water is being actively replenished by outgassing, and the production of H and $H_2$ is dependent on the $H_2O$ abundance. This leads to a slight decrease in H and $H_2$ mixing ratios, which further leads to the minor differences in the abundances of the other species.

Table 2 shows the escape velocities/fluxes, outgassing fluxes, and surface deposition velocities/fluxes for Cases 1 and 6 of the explicit flux model. The hydrogen escape velocities, which were derived using Case 1 and then applied to Case 6, are consistent with the diffusion-limited escape velocity of hydrogen at Earth's homopause (Hunten 1973). We found that greater escape velocities led to the depletion of hydrogen in the upper atmosphere with respect to the zero-flux results, while lower velocities led to excesses of H and $H_2$ without any changes in the shapes of their profiles. The hydrogen escape flux for Case 1 is consistent with that of Venus (Grinspoon 1993), which is unsurprising given their similar atmospheric hydrogen mole fractions (Miller-Ricci et al. 2009), while Case 6 has much lower escape fluxes, in accordance with its lower atmospheric hydrogen content. The same pattern holds for water outgassing: The flux for Case 1 is similar to that of Venus ($5.5 \times 10^5 - 10^7$ cm$^{-2}$ s$^{-1}$), while that of Case 6 is an order of magnitude lower than the given lower limit of the Venus water outgassing flux (Walker et al. 1970). By analogy this could mean that volcanism rates, and thus heat fluxes, are relatively high, but the water content of the lava is much lower. This is consistent with a desiccated upper mantle that lost its hydrogen during the early runaway greenhouse-induced magma ocean phase (Hamano et al. 2013; Luger & Barnes 2015).

The deposition of oxidants is crucial in determining the oxidative state of the atmosphere. For example, in the work of Zahnle et al. (2008), where $H_2O_2$ and $O_3$ are lost to aqueous reactions with ferrous iron and sulfides, the atmosphere tends to be converted entirely to CO. As $O_3$ and $H_2O_2$ are crucial to the catalytic cycles we have outlined, it is not surprising that losing them to the surface would result in high CO concentrations, as the efficiency of $CO_2$ regeneration would be vastly decreased. Zahnle et al. (2008) used an $O_3$ deposition velocity of 0.02 cm s$^{-1}$ for the surface of Mars, which is 3 and 8 orders of magnitude greater than that of our explicit flux results for Cases 1 and 6, respectively. Such low deposition velocities may be consistent with a surface much drier than that of Mars. Zahnle et al. (2008) used a surface relative humidity of water of 17% for their Mars model, which is 320 times higher than the relative humidity for our Case 1 results, and $8.5 \times 10^5$ times higher than that of our Case 6 results. Therefore, it is plausible that our deposition velocities are low due to the lack of an aqueous medium within which oxidation can proceed quickly. However, the exact relationship between $O_3$ deposition velocity and relative humidity is unknown, especially for very low water abundances. Alternatively, the planetary surface may be low in ferrous iron and sulfides due to the oxidation resulting from burial of hundreds of bars of oxygen during the aforementioned magma ocean phase.

Similar arguments can be made for $O_2$ deposition. Zahnle et al. (2008) attributed the surface flux of $O_2$ on Mars to UV photostimulated oxidation of magnetite in the presence of trace amounts of water. However, this may be insignificant in our case due to the low UV flux at our model planet's surface, as a result of our thicker atmosphere. Therefore, any $O_2$ flux at the surface is likely due to aqueous reactions with reducing species, as with $O_3$. An alternative simulation of the explicit flux cases using $O_2$ deposition instead of $O_3$ deposition yielded velocities of $3.7 \times 10^{-12}$ cm s$^{-1}$ and $3 \times 10^{-15}$ cm s$^{-1}$ for Cases 1 and 6, which are 5 and 8 orders of magnitude lower, respectively, than that assumed by Zahnle et al. (2008) for UV photostimulation ($4 \times 10^{-7}$ cm s$^{-1}$). Despite the different depositional mechanisms, these differences are similar in magnitude to that of our $O_3$ deposition results. Therefore, as both UV photostimulation and aqueous chemistry require some water, the reduced relative humidity at the surface of our desiccated planet may inhibit $O_2$ deposition by either mechanism just as it does $O_3$ deposition.

These calculations show that terrestrial planets that have experienced severe loss of water and oxidization of the crust and upper mantle, such as planets orbiting in the habitable zones of M dwarfs, can potentially possess an atmosphere similar to our zero-flux results provided that $CO_2$ is abundant in their atmospheres. Consequently, biosignature false positives may be prevalent for planets orbiting in the habitable zone of M dwarfs, and should be taken into consideration in future studies and observation campaigns, as we discuss below.

*4.2. Biosignature False Positives*

Our results show that $O_2$ mixing ratios of ~20% and $O_3$ mixing ratios >1 ppm are achievable on $H_2O$-depleted M dwarf terrestrial exoplanets due to photochemistry. These values are similar to that of modern Earth (Yung & DeMore 1999), where the abundant $O_2$ and $O_3$ are produced by biological photosynthesis. In this section we calculate their spectral signatures and consider their impact as potential false positive biosignatures.

Figure 8 shows the thermal emission spectrum of our model atmosphere for Cases 1 (red), 3 (orange), and 6 (green), as generated by the Spectral Mapping Atmospheric Radiative Model (SMART) (Meadows & Crisp 1996). We also plot the spectrum of Earth (black) generated by the extensively validated Virtual Planetary Laboratory (VPL) 3D spectral Earth model (Robinson et al. 2010; Robinson et al. 2011; Robinson et al. 2014), divided by two for comparison. We see immediately that not only are our three cases difficult to separate, but all three cases are very different from Earth's atmosphere as seen in the mid-IR. This is mostly due to the completely different temperature profile assumed and the lack of water vapor, which absorbs at 6.3 and beyond 18 μm, though there is a slight decrease in flux at 6.3 μm for our wettest case (Case 1). Our model atmosphere also lacks methane, which absorbs at 7.7 μm. All spectra feature a deep 15-μm $CO_2$ band, though it is deeper for our model atmosphere due to the cooler stratosphere. Finally, there is weak $O_3$ absorption at 9.7 μm that gets progressively stronger as our model atmosphere becomes drier, but it never approaches the strength of $O_3$ absorption in Earth's atmosphere. The weakness of this absorption is caused by the emission occurring near the surface, such that the brightness temperature of the $O_3$ absorption is similar to the surface temperature, which sets the continuum on either side of the absorption feature. Thus, from the thermal emission there is a clear distinction between actual Earth-like planets and those with abundant atmospheric $O_2$ and $O_3$ caused by low water content and photolysis.

Figure 9 shows the normalized reflectance of the same cases as in Figure 8 compared to the reflectance of Earth, again generated by SMART and the VPL Earth model, respectively. A grey surface albedo of 0.15 is assumed for our model planet. We see that the $O_3$ in our model atmosphere can be detected via deep $O_3$ absorption in the NUV similar in strength to $O_3$ absorption in Earth's atmosphere, and our driest case (Case 6) even features a deeper Chappuis band (~0.6 μm) than Earth. However, whereas all of our cases exhibit a smooth fall-off in reflectance past 0.7 μm, punctuated by $CO_2$ absorption, Earth's reflectance in this region is dominated by water absorption, and thus water again acts as the discriminating species between Earth and our model planet.

From these comparisons it is clear that the condition necessary to produce abundant $O_2$ and $O_3$ in $CO_2$-dominated atmospheres – low water content – leads to a clear distinction in planetary spectra from those of Earth-like planets via the disappearance of strong water vapor absorption bands. Thus, biologically active, Earth-like planets can be distinguished from the planetary environments presented here using the 0.95 μm water vapor band, the numerous water bands throughout the near infrared, or, in the thermal infrared, the 6.3 μm water band, which are all observable by JWST (Beichman et al. 2014). However, caution must be exercised, as high-altitude cloud cover could potentially mask spectral signatures of water vapor.

*4.3. Atmospheric Evolution*

In our calculations we fix the pressure, temperature, and eddy diffusion coefficient profiles. As we have shown, considerable atmospheric evolution occurs over the 10 Gyr duration of each simulation, including the formation of a significant $O_3$ layer comparable to that of Earth and the conversion of ~40% of the atmospheric $CO_2$ into CO and $O_2$. Furthermore, the liberation of CO and O from $CO_2$ increases the total number of molecules in the atmosphere, but this should be balanced by the decrease of the molecular weight of the atmosphere from ~44 g mol$^{-1}$ to ~37 g mol$^{-1}$, thereby increasing the scale height such that local number densities should remain constant. The $O_3$ layer in the stratosphere should raise the temperature such that a temperature inversion results due to heating by photolysis, much like in Earth's stratosphere (Park & London 1974). The decrease in $CO_2$ should result in warming of the stratosphere and mesosphere due to lower rates of radiative cooling (Roble & Dickinson 1989; Brasseur & Solomon 2005), as well as a cooling of the troposphere due to decreased greenhouse effects. Finally, the increase in $O_2$ should result in heating by photolysis in the mesosphere and thermosphere (Park & London 1974). In all, the temperature profile should increase above the tropopause and evolve in shape towards that of Earth, while below the tropopause the temperature will likely decrease.

The reaction rates will change in response to these temperature variations, as most of the reaction rate constants are exponential functions of inverse temperature (Table 1). Of the chemical reactions considered here, R2 (low pressure limit), R3, and R7 have rate constants that are inversely proportional to temperature, while R2 (high pressure limit), R4, R5, and R8 have rate constants that are directly proportional to temperature. Of the photolysis reactions, R1 decreases for decreasing temperatures (Schmidt et al. 2013) and R6 varies negligibly with temperature (Nicovich & Wine 1988). Photolysis of $O_2$ and $O_3$ (R9) decrease slightly with decreasing temperatures (Hudson et al. 1966; Lean and Blake 1981).

The combination of the temperature changes and the response of the reaction rates seem to suggest that (1) $CO_2$ destruction will increase as temperature increases in the stratosphere, (2) $CO_2$ regeneration via S2 and S3 will decrease due to decreasing temperatures in the troposphere, and (3) $CO_2$ regeneration via R5 will also decrease, again due to the cooler troposphere. In other words, the atmosphere may lose much more $CO_2$ if the temperature evolution were taken into account. Quantitatively verifying these hypotheses require running coupled photochemistry and radiative transfer models, which is beyond the scope of this study.

*4.4. Condensation*

For many of our cases, $H_2O$ is supersaturated in the stratosphere/mesosphere by the end of the 10 Gyr model run due to equilibration among the different hydrogen species. The formation of ice clouds would affect the UV radiation environment in the lower atmosphere, but whether the UV is enhanced or attenuated depends on location above or below the clouds, the solar zenith angle, cloud particle properties and distribution, and cloud optical depth. Typical variations in UV intensity on Earth's surface due to clouds are on the order of 25% in either direction while enhancement tends to occur above the clouds (Bais et al. 2006). Given the degree of supersaturation in our wetter runs, however, attenuation is more likely than enhancement below the clouds. This would decrease the $H_2O_2$ and $O_3$ photolysis rates and cause an increase in the number densities of these species, which in turn would decrease the efficiency of S2 and R5 while increasing the efficiency of S3. The magnitude to which these pathways change will depend on how much UV is attenuated as a function of wavelength, as well as the variability time scale of the clouds – if this time scale is sufficiently short, it may be irrelevant for the long time scales considered here.

The existence of water ice clouds may also result in heterogeneous chemistry that impacts the stability of $CO_2$ (Atreya & Blamont 1990). For example, observations of high $O_3$ mixing ratios over the northern polar region of Mars during northern spring is likely due to the uptake of $HO_x$ radicals by cloud particles (Lefèvre et al. 2008). If similar processes occurred in our model atmosphere, then the decrease in $HO_x$ species would reduce the efficiency of S2 and S3, while the buildup in $O_3$ would increase the efficiency of S3 and R5. As S2 is responsible for the highest rate of $CO_2$ regeneration, inclusion of heterogeneous chemistry will likely result in reduced amounts of $CO_2$, even when the atmospheric hydrogen content is relatively high. This is consistent with simulations of the Martian atmosphere that included heterogeneous chemistry (Atreya & Gu 1994; Lefèvre et al. 2008). The magnitude of these effects will depend on the reactive surface area of the cloud particles and the amount of condensable water vapor in the atmosphere.

In drier cases, the lack of significant cloud formation and rainout may produce a substantial silicate dust layer, as in the near-surface atmosphere of Mars (Hamilton et al. 2005). These particles may also participate in heterogeneous chemistry and UV attenuation (Atreya & Gu 1994), though the magnitude of the effect is likely less than that of water ice particles (Nair et al. 1994).

## 5. SUMMARY AND CONCLUSIONS

We have investigated the stability of $H_2O$-depleted, $CO_2$-dominated atmospheres of terrestrial exoplanets orbiting near the outer edge of the habitable zones of M dwarfs. These atmospheres may be common due to the desiccation and oxidation of the outer layers of these planets during the host stars' pre-main sequence, when it is much more luminous than later in its evolution. We found that these atmospheres are stabilized against photolytic conversion to CO by a series of catalytic cycles and reactions. On Mars, $CO_2$ is maintained through the generation of free hydrogen via water photolysis leading to the oxidation of CO by $HO_x$ species. By comparison, for a planet orbiting an M dwarf with severe water depletion, $H_2O_2$ photolysis dominates as the main driver of $CO_2$ regeneration due to the lower NUV fluxes of M dwarfs leading to a buildup of $H_2O_2$. This cycle is capable of maintaining 80 – 90% of the original $CO_2$ content of the atmosphere. However, for atmospheric hydrogen content <1 ppm, $H_2O_2$ and $HO_2$ concentrations decrease and the $H_2O_2$ cycle (S2) breaks down, resulting in decreasing $CO_2$ and increasing $O_2$ and CO. The rise in $O_2$ leads to higher $O_3$ mixing ratios, which can now participate in the $CO_2$ regeneration process. Due to the low $HO_2$ concentrations, the $O_3$ cycle (S3) can only maintain half of the $CO_2$ in the original atmosphere. If the atmospheric hydrogen content is <0.1 ppm, any reactions involving H-species such as H or $HO_2$ become irrelevant, and the spin-forbidden recombination reaction R5 dominates. This reaction is actually more effective than S3, as $O_3$ (the source of O) increases in abundance as H decreases, and thus only 40% of the atmospheric $CO_2$ is converted to CO and $O_2$. These values may be subject to change given different atmospheric temperatures or surface pressures, but these catalytic cycles and reactions are likely general due to their self-limiting nature, i.e. if one catalytic cycle is inefficient, then another will take over.

Our model produces abundances of abiotic $O_2$ and $O_3$ comparable to that produced by photosynthetic organisms on modern Earth. This is due to our assumption of a low-$H_2O$ atmosphere and dry surface, which decreases the $CO_2$ regeneration rate and the surface deposition rate of oxidizing species. These conditions may be applicable to terrestrial exoplanets in the habitable zones of M dwarfs (Luger & Barnes 2015). The high mixing ratios of $O_2$ and $O_3$ can potentially pose as false positives in the search for biosignatures, but a simultaneous search for water vapor could potentially differentiate between an actual Earth-like planet with biologically mediated atmospheric disequilibrium from a water-depleted planet with atmospheric disequilibrium caused by photochemistry. However, cloud cover can dilute this effect by decreasing the strength of water features. Meanwhile, if surface deposition were significant, then the catalytic cycles would break down due to loss of $O_3$ to the surface resulting in low $O_2$ and $O_3$ abundances. This would lead to much greater losses of $CO_2$ and ultimately the formation of a CO atmosphere.

The sensitivity of these results to temperature changes and clouds is highly complex and nonlinear, owing to the unknowns in UV-cloud radiative transfer, heterogeneous chemistry, and the feedback between temperature and reaction rates. A coupled photochemical and radiative transfer model may be necessary to fully elucidate the evolution of these exotic atmospheres.

**ACKNOWLEDGEMENTS**

We thank K. Willacy, M. Allen, and R. L. Shia for assistance with the setting up and running of the KinetgenX code. We thank V. Meadows and R. Barnes for their valuable inputs. This research was supported in part by the Venus Express program via NASA NNX10AP80G grant to the California Institute of Technology, and in part by an NAI Virtual Planetary Laboratory grant from the University of Washington to the Jet Propulsion Laboratory and California Institute of Technology. Support for RH's work was provided in part by NASA through Hubble Fellowship grant #51332 awarded by the Space Telescope Science Institute, which is operated by the Association of Universities for Research in Astronomy, Inc., for NASA, under contract NAS 5-26555. Part of the research was carried out at the Jet Propulsion Laboratory, California Institute of Technology, under a contract with the National Aeronautics and Space Administration.

**FIGURES**

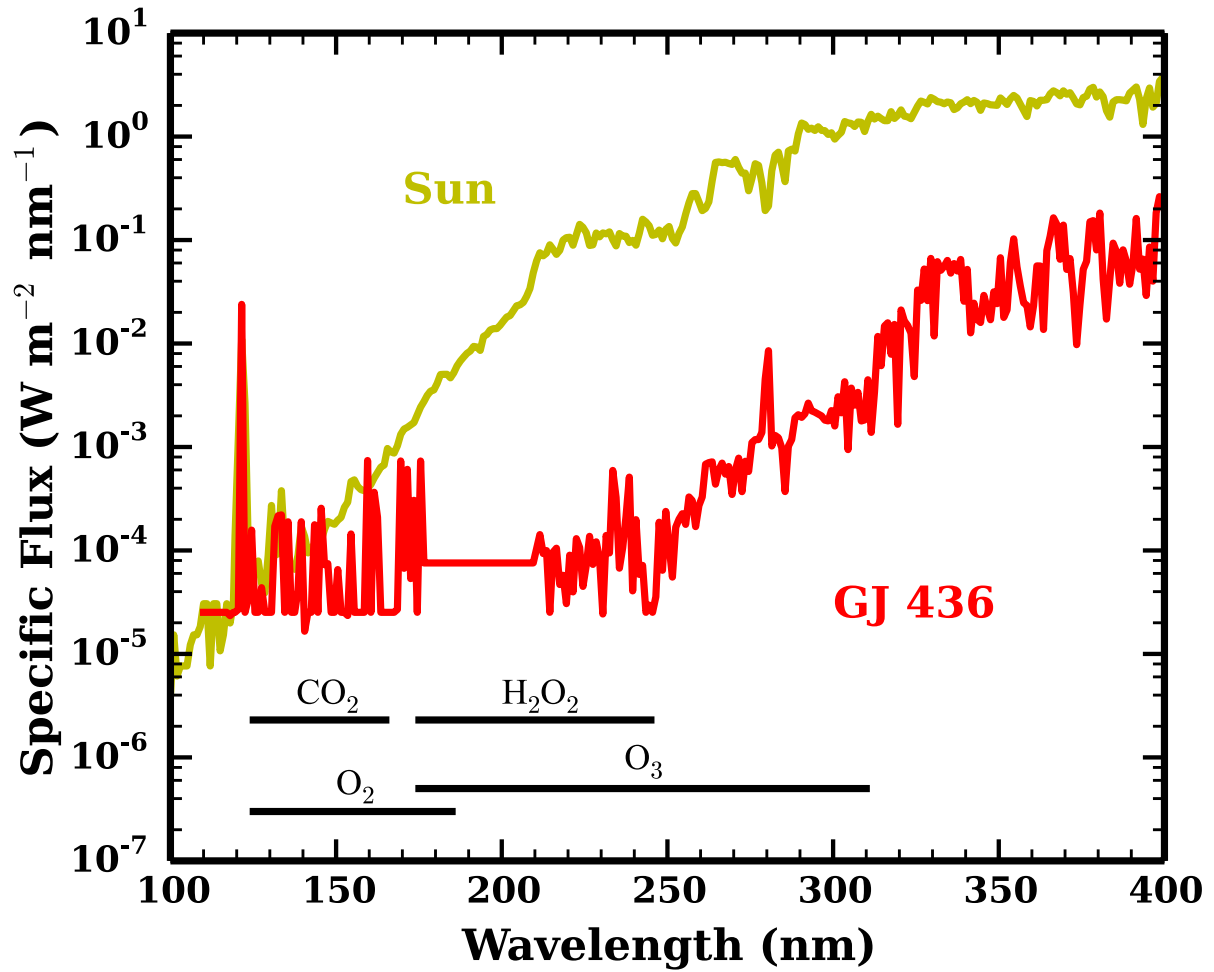

Figure 1. The spectra of the Sun (WMO 1985) (yellow) and GJ 436 (France et al. 2013) (red) scaled such that the total flux of each is identical to the total flux received by Earth at 1AU (~ 1360 W m$^{-2}$). The wavelengths at which photolysis of $CO_2$, $O_2$, $O_3$, and $H_2O_2$ is most efficient are added for comparison (Tian et al. 2014).

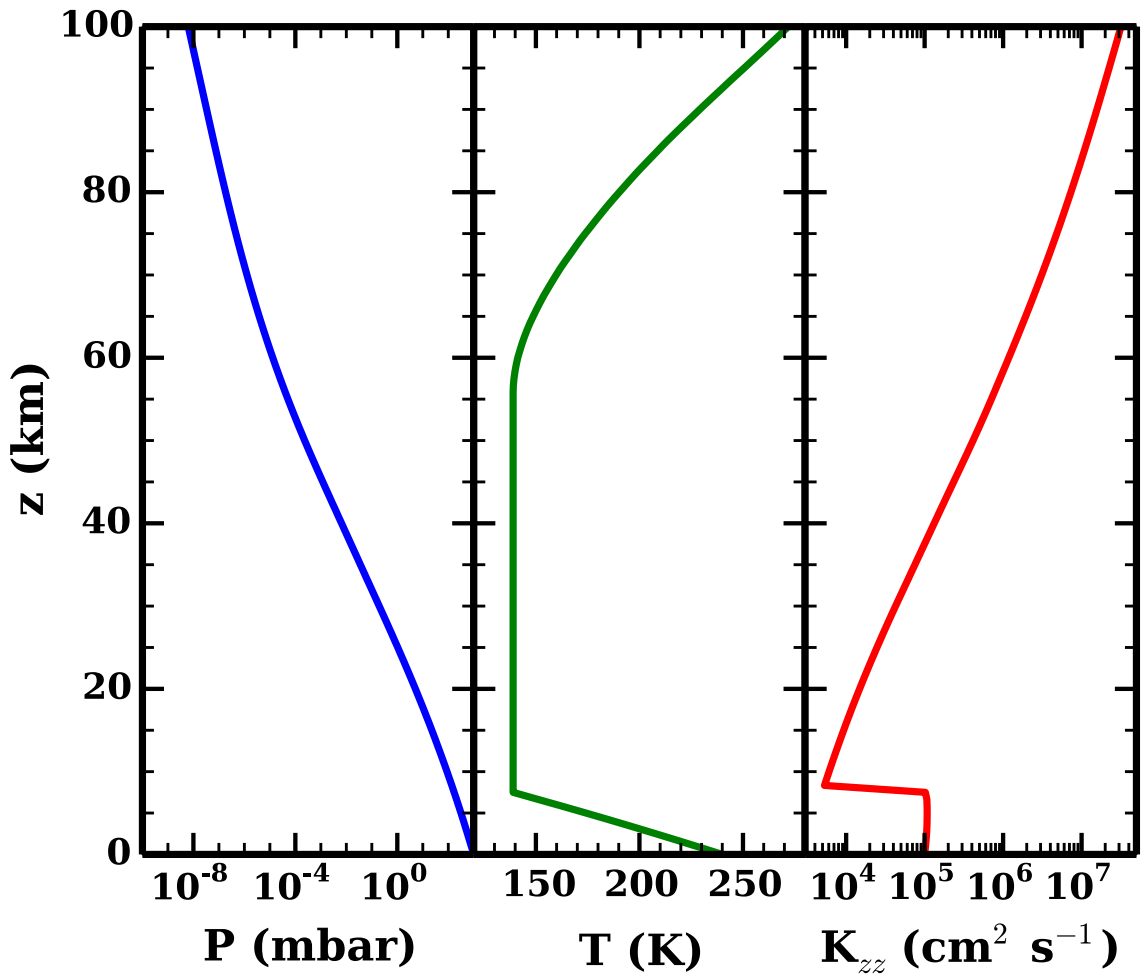

Figure 2. Model atmospheric pressure P (left), temperature T (center), and eddy diffusion coefficient $K_{zz}$ (right) as functions of altitude z.

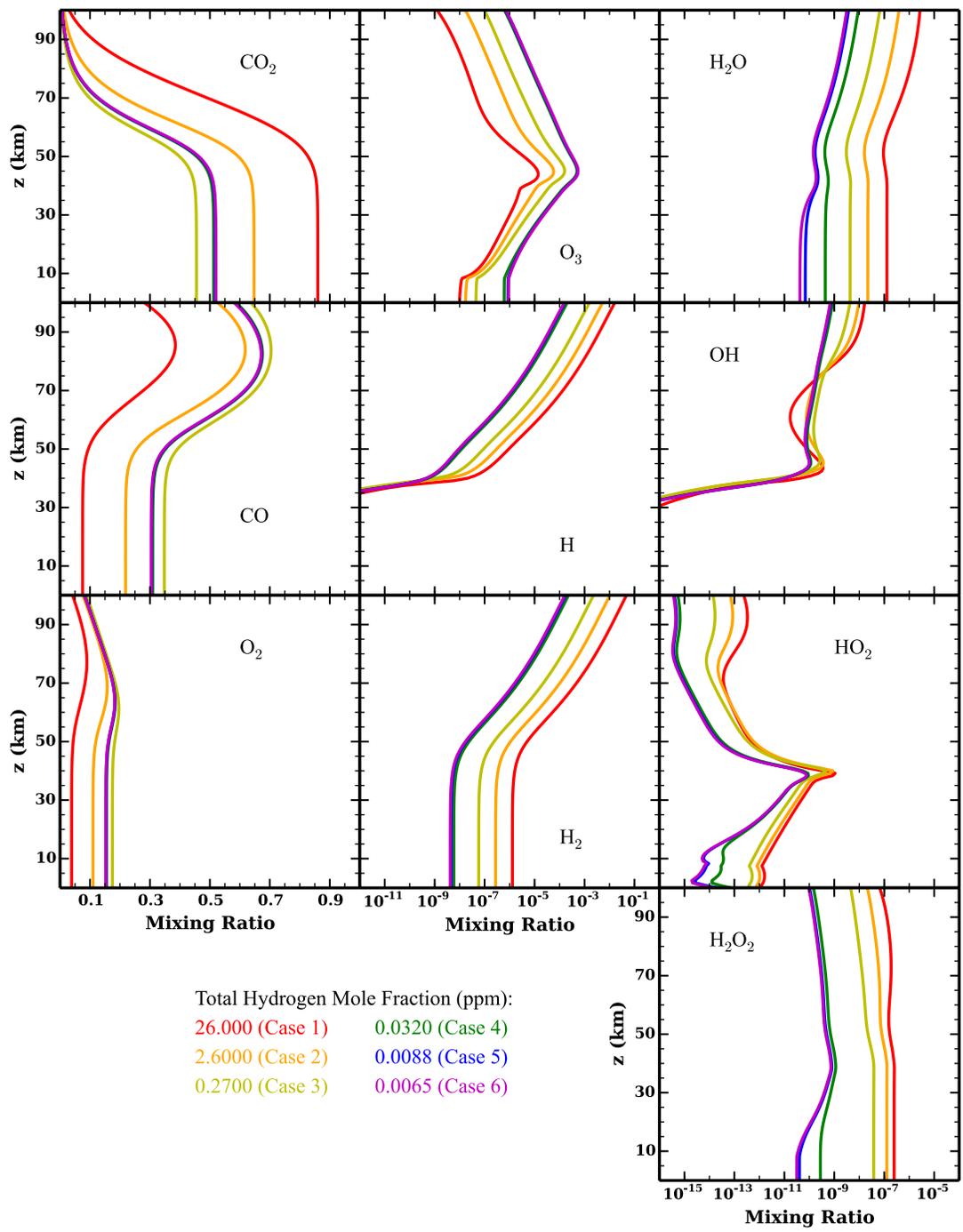

Figure 3. Mixing ratio profiles of $CO_2$, CO, $O_2$, $O_3$, H, $H_2$, $H_2O$, OH, $HO_2$, and $H_2O_2$ for Cases 1 (red), 2 (orange), 3 (yellow), 4 (green), 5 (blue), and 6 (magenta). Note the different x-axis

values for the three different columns. For many of the species, the curves for cases 4, 5, and 6 overlap each other.

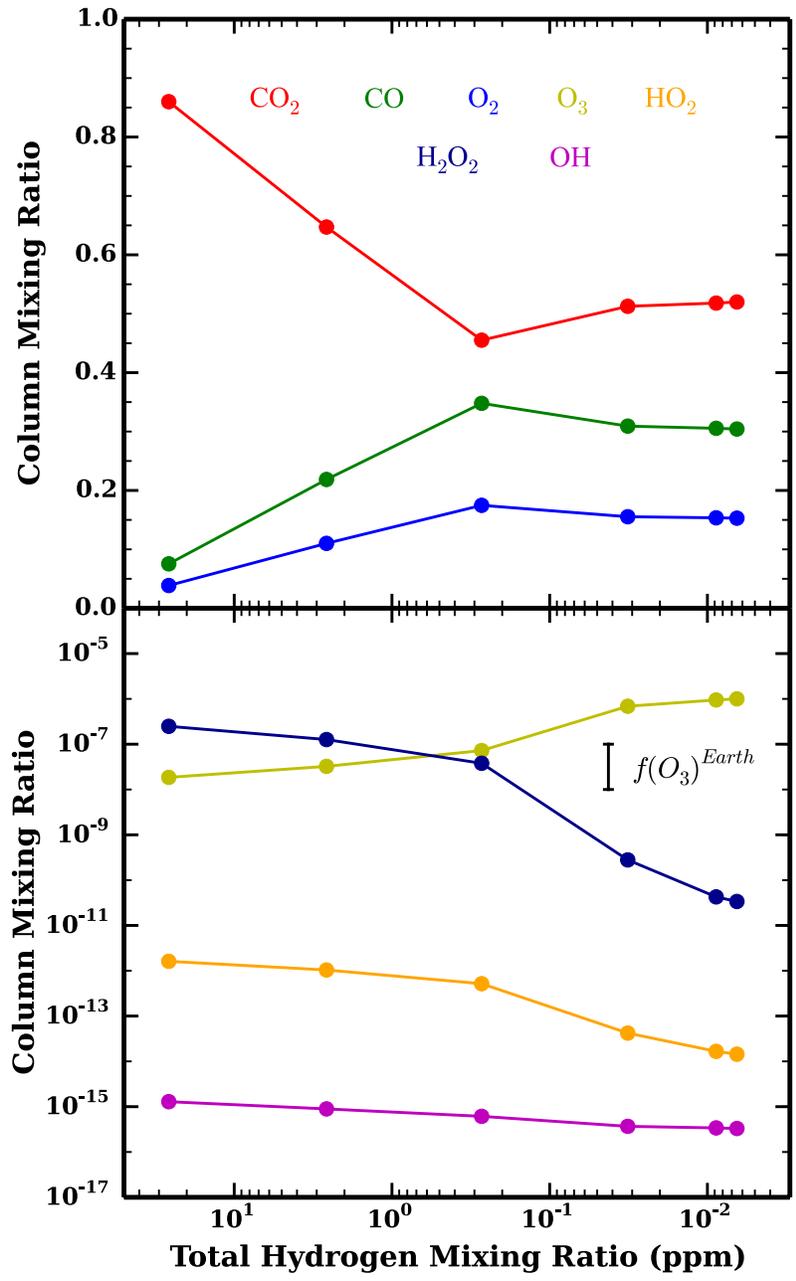

Figure 4. Column-integrated mixing ratios of (top) $CO_2$ (red), CO (green), $O_2$ (blue), (bottom) $O_3$ (yellow), $HO_2$ (orange), $H_2O_2$ (dark blue), and OH (magenta) as functions of the total atmospheric hydrogen mole fraction for Cases 1 through 6. This mole fraction is calculated by dividing the number of hydrogen atoms in the atmosphere by the total number of atoms in the atmosphere. The points indicate the results of the actual cases, which are connected by lines. In the bottom panel we show the range in Earth's $O_3$ mixing ratio $f(O_3)^{Earth}$ (Yung & DeMore 1999,

pp. 318). Note the different y-axis scales of the top and bottom panels.

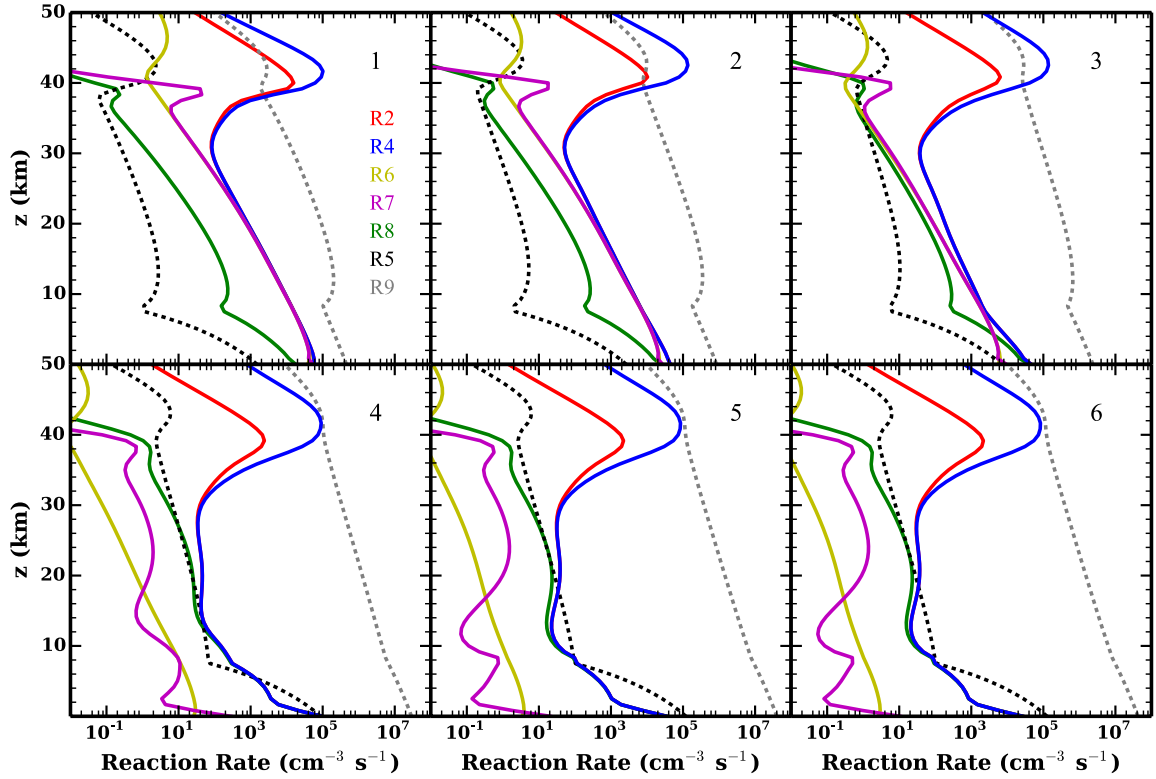

Figure 5. Reaction rates of R2: $H + O_2 + M \rightarrow HO_2 + M$ (red), R4: $OH + CO \rightarrow CO_2 + H$ (blue), R5: $CO + O + M \rightarrow CO_2 + M$ (black, dotted line), R6: $H_2O_2 + h\nu \rightarrow 2OH$ (yellow), R7: $2HO_2 \rightarrow H_2O_2 + O_2$ (magenta), R8: $HO_2 + O_3 \rightarrow 2O_2 + OH$ (green), and R9: $O_3 + h\nu \rightarrow O_2 + O$ (gray, dotted line) as functions of altitude in the lower atmosphere for (1) Case 1, $f(H)_{tot} \sim 26$ ppm; (2) Case 2, $f(H)_{tot} \sim 2.6$ ppm; (3) Case 3, $f(H)_{tot} \sim 0.27$ ppm; (4) Case 4, $f(H)_{tot} \sim 0.032$ ppm; (5) Case 5, $f(H)_{tot} \sim 0.0088$ ppm; and (6) Case 6, $f(H)_{tot} \sim 0.0065$ ppm.

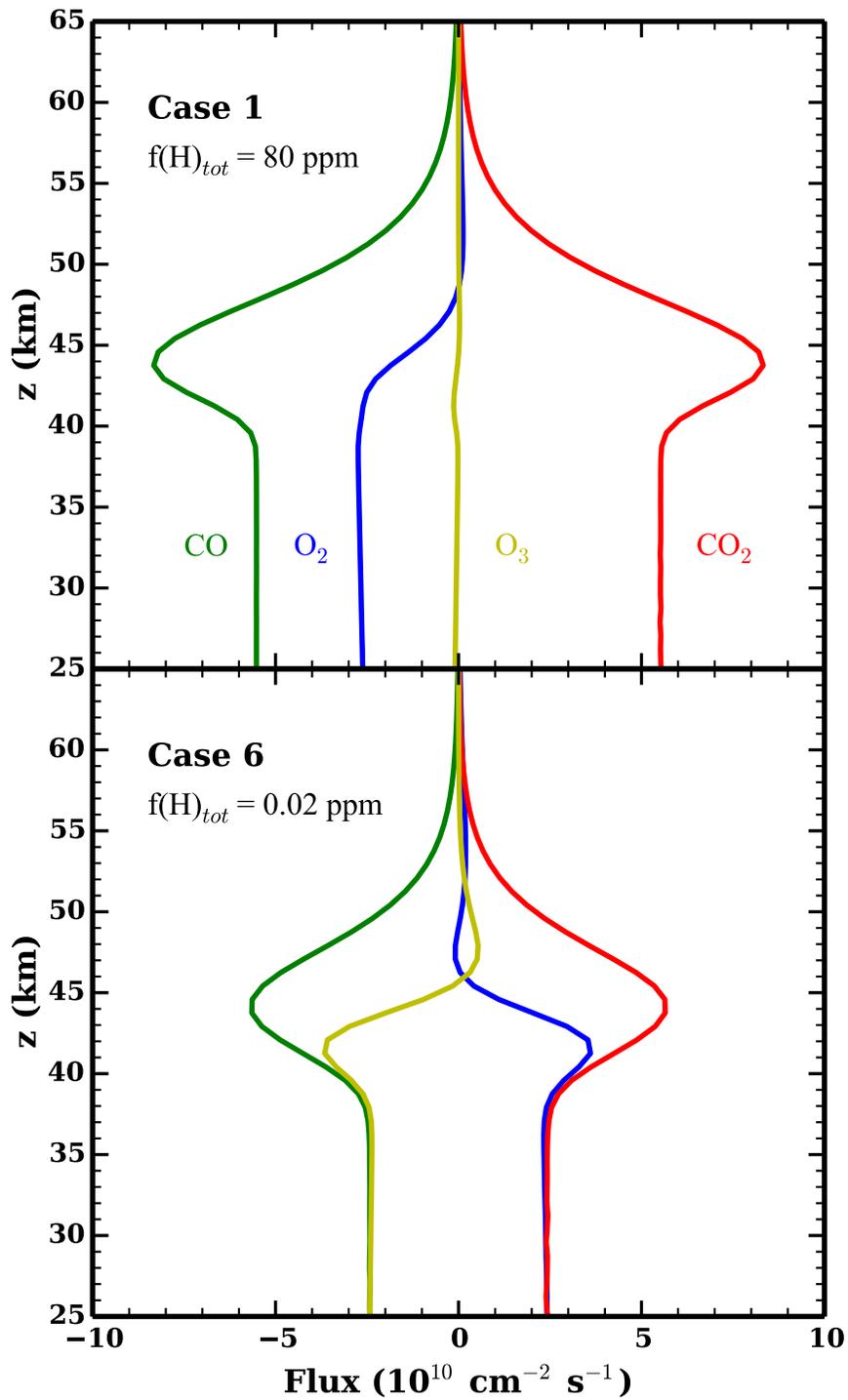

Figure 6. Vertical fluxes of $O_2$ (blue), $O_3$ (yellow), $CO_2$ (red), and CO (green) in the lower atmosphere for Cases 1 (top) and 6 (bottom). Upward fluxes are positive while downward fluxes are negative.

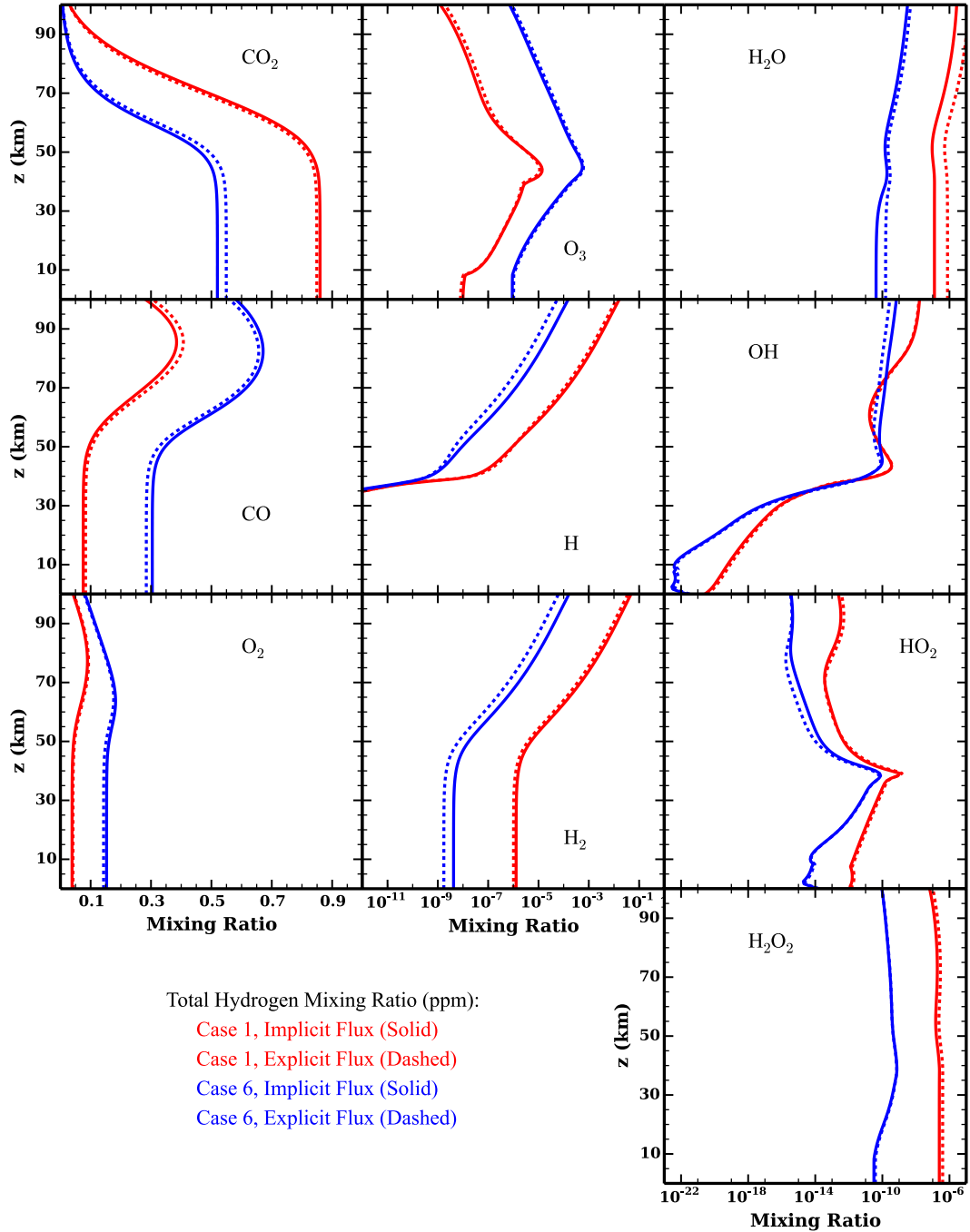

Figure 7. Mixing ratio profiles of $CO_2$, CO, $O_2$, $O_3$, H, $H_2$, $H_2O$, OH, $HO_2$, and $H_2O_2$ for Case 1 of the nominal, zero-flux model (red, solid), Case 1 of the explicit flux model (red, dashed), Case 6 of the nominal model (blue, solid), and Case 6 of the explicit flux model (blue, dashed). Note the different x-axis values for the three different columns.

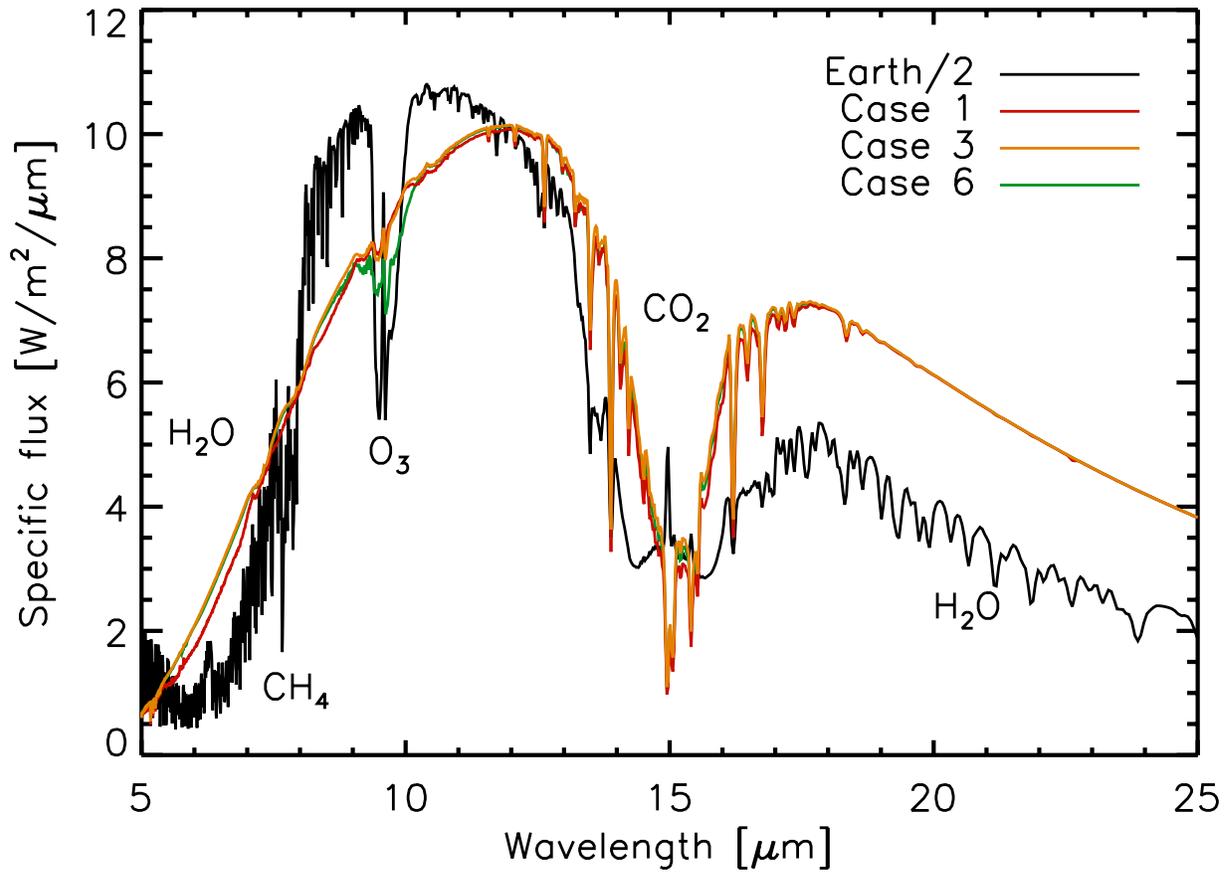

Figure 8. The thermal emission spectra of our model atmosphere for cases 1 (red), 3 (orange), and 6 (green) in the mid-IR compared to that of Earth (black) (Robinson et al. 2011), divided by 2 for comparison.

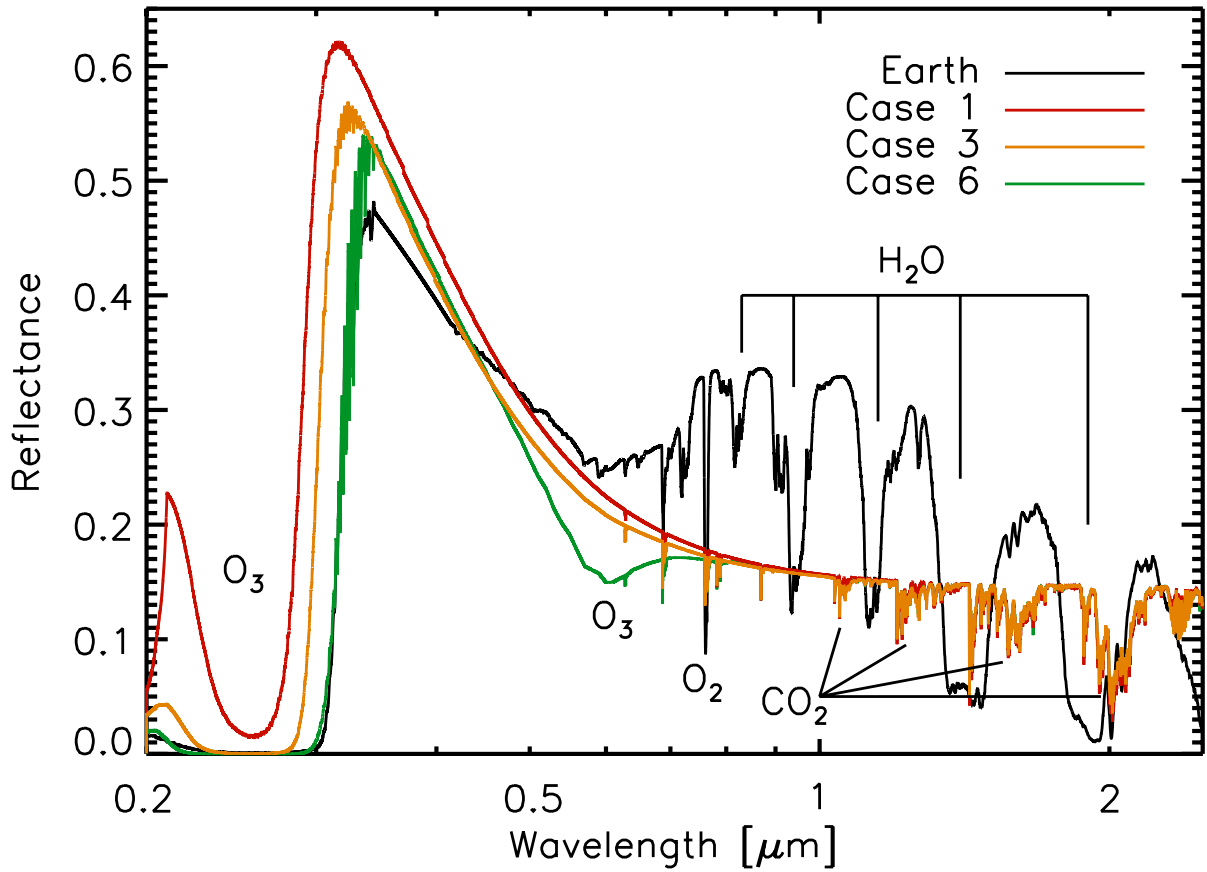

Figure 9. The normalized reflectance of our model atmosphere for cases 1 (red), 3 (orange), and 6 (green) from the near-UV to the near-IR compared to that of Earth (black) (Robinson et al. 2011). A grey surface albedo of 0.15 is assumed for our model planet.

Table 1: Important reactions involved in $CO_2$ destruction and regeneration. The rate constants are updated from those of Nair et al. (1994) using Sander et al. (2011) and references therein.

|    | Reaction | Rate Constant |
|----|----------|---------------|
| R1 | $CO_2 + h\nu \rightarrow CO + O$ | $1.025 \times 10^{-8}$ |
| R2 | $H + O_2 + M \rightarrow HO_2 + M$ | $k_0 = 7.3 \times 10^{-29} T^{-1.3}$ <br> $k_\infty = 2.4 \times 10^{-11} T^{0.2}$ |
| R3 | $O + HO_2 \rightarrow OH + O_2$ | $3.0 \times 10^{-11} e^{200/T}$ |
| R4 | $OH + CO \rightarrow CO_2 + H$ | $4.9 \times 10^{-15} T^{0.6}$ |
| R5 | $CO + O + M \rightarrow CO_2 + M$ | $1.6 \times 10^{-32} e^{-2184/T}$ |
| R6 | $H_2O_2 + h\nu \rightarrow 2OH$ | $1.273 \times 10^{-6}$ |
| R7 | $2HO_2 \rightarrow H_2O_2 + O_2$ | $3.0 \times 10^{-13} e^{460/T}$ |
| R8 | $HO_2 + O_3 \rightarrow 2O_2 + OH$ | $1.0 \times 10^{-14} e^{-490/T}$ |
| R9 | $O_3 + h\nu \rightarrow O_2 + O$ | $2.898 \times 10^{-6}$ |

Notes: Rate constants for photolysis reactions are for the top of the atmosphere. Units for photochemical, two-body, and three-body reactions are $s^{-1}$, $cm^3\ s^{-1}$, and $cm^6\ s^{-1}$, respectively. $k_0$ and $k_\infty$ are the low- and high-pressure limit rate constants, respectively.

Table 2: Boundary conditions for the explicit flux model that reproduced the corresponding cases for the zero-flux model. All velocities are given in units of cm s$^{-1}$ and all fluxes are given in units of cm$^{-2}$ s$^{-1}$.

|  | Case 1 | Case 6 |
|---|---|---|
| H escape velocity / flux | 5.5 / 1.2 × 10$^7$ | 5.5 / 6 × 10$^4$ |
| H$_2$ escape velocity / flux | 0.3 / 2 × 10$^6$ | 0.3 / 3 × 10$^3$ |
| H$_2$O outgassing flux | 8 × 10$^6$ | 3.3 × 10$^4$ |
| O$_3$ deposition velocity / flux | 10$^{-5}$ / 2.7× 10$^6$ | 3 × 10$^{-10}$ / 1.1 × 10$^4$ |